\documentclass[runningheads]{llncs}
\usepackage{graphicx}
\usepackage{kotex}
\usepackage{cite}
\usepackage{multirow}
\usepackage{array}
\usepackage{hhline}
\newcolumntype{L}[1]{>{\raggedright\let\newline\\\arraybackslash\hspace{0pt}}m{#1}}
\newcolumntype{C}[1]{>{\centering\let\newline\\\arraybackslash\hspace{0pt}}m{#1}}
\newcolumntype{R}[1]{>{\raggedleft\let\newline\\\arraybackslash\hspace{0pt}}m{#1}}

\begin{document}
\title{Attention mechanisms for physiological signal deep learning: which attention should we take?}


\titlerunning{Attention mechanisms for physiological signal deep learning}

\author{
Seong-A Park\inst{1} 
\and
Hyung-Chul Lee\inst{1,2} 
\and
Chul-Woo Jung\inst{1,2} 
\and
Hyun-Lim Yang\inst{1,}\thanks{Corresponding author.} 
}
\authorrunning{Park et al.}


%
\institute{Department of Anesthesiology and Pain Medicine, \\Seoul National University Hospital, Seoul, Republic of Korea \and
Department of Anesthesiology and Pain Medicine, Seoul National University College of Medicine, Seoul, Republic of Korea \\
\email{hlyang@snu.ac.kr}
}

\maketitle 
\begin{abstract}
Attention mechanisms are widely used to dramatically improve deep learning model performance in various fields.
However, their general ability to improve the performance of physiological signal deep learning model is immature. 
In this study, we experimentally analyze four attention mechanisms (e.g., squeeze-and-excitation, non-local, convolutional block attention module, and multi-head self-attention) and three convolutional neural network (CNN) architectures (e.g., VGG, ResNet, and Inception) for two representative physiological signal prediction tasks: the classification for predicting hypotension and the regression for predicting cardiac output (CO). 
We evaluated multiple combinations for performance and convergence of physiological signal deep learning model. 
Accordingly, the CNN models with the \textit{spatial} attention mechanism showed the best performance in the classification problem, whereas the \textit{channel} attention mechanism achieved the lowest error in the regression problem. 
Moreover, the performance and convergence of the CNN models with attention mechanisms were better than stand-alone self-attention models in both problems. 
Hence, we verified that convolutional operation and attention mechanisms are complementary and provide faster convergence time, despite the stand-alone self-attention models requiring fewer parameters.
\end{abstract}
\section{Introduction}
Deep learning has dramatically improved the predictability of various phenomena based on input data of past events. 
For natural language processing, recurrent neural networks (RNNs) are particularly effective in analyzing time-series sequences \cite{lstm, bilstm, gru}. 
For image processing, convolutional neural networks (CNNs) that mimic human visual cognitive functions have grown in popularity \cite{vggnet, resnet, inception}.  
However, both methods have shortcomings, such as the RNN’s vanishing gradient and information loss problems \cite{rnn_gradient_vanishing}, which limits performance, and the CNN’s locality of pixel dependency \cite{alexnet}, which make it goes deeper. 
To overcome these roadblocks, attention mechanisms have been used to enable neural models to pay closer attention to the most important parts of the data while ignoring irrelevant parts \cite{AttentionCVsurvey}.
It gives higher weight to parts that are more relevant to produce output, and lower weights to parts that are not. 
Bahadnau et al.\cite{attention_first} introduced this idea to machine translation, resulting in superior performance over canonical RNNs. 
Similar concept of attention mechanism was also introduced, e.g., Luong et al. \cite{attention_luong}, and the other types of attention mechanisms were also proffered which tailored to computer vision applications \cite{nonlocal, senet, cbam}.

In recent days, self-attention-based mechanisms had been replaced the canonical deep learning architectures and are positioned as a mainstream of AI research.
Vaswani et al. \cite{transformer} proposed a deep learning model that skipped the RNN and applied a self-attention mechanism by itself (so-called Transformer), achieving superior performance in machine translation and document generation. 
Dosoviskiy et al.\cite{VIT} proposed a vision transformer, which a variant of the Transformer for image classification tasks, outperforming canonical CNNs with substantially fewer computations. 
Subsequently, self-attention-based deep learning was used to predict protein structures \cite{alphafold2}, compiler graph optimizers \cite{MLforSystem}, and audio generation methods \cite{MusicTransformer}. 

Consequently, the application of deep learning to physiological signal analysis has been considered \cite{physiological_signal_review}.
For example, Hannun et al. \cite{ecg_ng} built a CNN that detects arrhythmia from electrocardiogram (ECG), showing human expert-level performance. 
As in other domains, attention mechanisms have been used to improve performance in physiological signal analysis. 
Mousavi et al. \cite{sleepEEGnet} proposed an attention-based CNN+RNN network to predict sleep stages from single-channel electroencephalogram. 
Yang et al. \cite{adult_apco} built a CNN with attention blocks to predict stroke volume from arterial blood-pressure waveform. 
Unfortunately, all of these methods were tuned for specific signal types or tasks, and the best attention mechanisms for general field use for physiological signal analysis was not determined. 

In this study, we experimentally determine which CNN architectures and attention mechanisms are the best for analyzing physiological signals. 
We focus on attention mechanisms used in computer vision, as the various features of physiological signal processes are similar, and the challenges of accurately predicting and classifying the presence of signal and object anomalies are closely related. 
Hence, we considered the three types of CNN models which popular for image processing and four types of attention mechanisms which suggested for computer vision tasks.
Notably, a physiological signal generally has a smaller dimension than does an image, and the attention mechanism designed for computer vision may reduce efficiency by adding unnecessary calculations. 
Additionally, in a computer vision problem, discriminating feature detection is the main task, whereas in physiological signal analysis, not only is detecting discriminating features important, but detecting signal trends is also crucial. 
Therefore, for effective and efficient use of attention mechanisms, it is necessary to analyze how each attention mechanisms affects physiological signal analysis. 
To the best of our knowledge, this study is the first attempt to identify the most effective attention mechanism for physiological signal analysis using deep learning. 
We believe that our work will enable generalizable physiological signal deep learning, including the development of prototypes.

\section{Methods}
In this study, we analyze the efficacy of three CNN architectures (e.g., VGG-16 \cite{vggnet}, ResNet-18 \cite{resnet}, and Inception-V1 \cite{inception}) with four types of attention mechanisms (e.g., squeeze-and-excitation (SE) \cite{senet}, non-local (NL) \cite{nonlocal}, convolutional block attention module (CBAM) \cite{cbam}, and multi-head self-attention (MSA) \cite{transformer}) for physiological signal deep learning. 
Each model uses unique feature extraction modules. 
VGG module includes two or three consecutive convolution layers and a pooling layer. 
ResNet module contains two consecutive convolution layers and a residual path. 
Inception module includes three convolution layers and a pooling layer in parallel. 
The CNN models used in this study are tailored to modality and dimension differences between image and physiological signal data. 
Detailed reduction criteria are described in Appendix.

The SE module is a \textit{channel} attention mechanism. 
It encodes features with a squeeze part and decodes it with an excitation part to increase the quality of feature representation by considering the interdependency of channel information. 
The NL module is a \textit{spatial} attention mechanism that calculates global feature information with covariance-like self-attention, which can overcome the locality of pixel dependency of CNN model, in which they fail to extract relational features between the first and last points of the input segment.
CBAM is a \textit{channel + spatial} attention mechanism.
It performs channel-wise attention which is similar to SE module and performs spatial attention mechanism in that it sequentially reduces the feature size using multiple pooling and convolutional layers.
The MSA module \cite{transformer} is a stand-alone spatial self-attention method comprising multiple scaled dot-product attention layers in parallel, which use input data itself for queries, keys, and values. 
It analyzes how the given input data are self-related and helps extract enriched feature representations. 
The first three attention modules are harmonized to CNN models, but MSA does not use intermediate convolutional layers. 
A total of 13 types deep learning models (i.e., three pure CNN-based models, nine attention involved CNN-based models, and an MSA-based model) are compared.

Each model is trained to solve two representative physiological signal problems: classification for predicting intraoperative hypotension and regression for predicting intraoperative cardiac output (CO). 
Unexpected hypotension is a critical event that requires prompt intervention. 
Many risk factors have been revealed, but they do not help reduce its incidence or duration. 
Therefore, early prediction and prevention are crucial. 
Several studies have attempted to predict hypotension using deep learning \cite{hypotension, hypotension_ml}. 
We followed their methods of predicting hypotension events within 5 min of occurrence.

ECG, plethysmography (PPG), and demographic data were used as input variables for classification task. 
The output variable was binary, the positive label was defined as hypotension (mean arterial blood pressure$\leq$65 mmHg) lasting$>$1 min, and the negative label for otherwise.
A pair of 20-s input segments of ECG and PPG waveforms and demographic data were extracted to predict events within 5-min. 
For preprocessing, we removed segments with ECG outside a range of $-$2 to 4.5mV or a PPG range of zero (unitless) or less.

CO, the volume of blood being pumped by the heart per minute, is used to monitor and optimize systemic oxygen and drug delivery in critically ill or high-risk surgical patients. 
Especially for surgical patients, it is directly related to postoperative complications; hence, immediate treatment to keep CO levels between 4 and 8L/min during surgery may improve patient outcomes \cite{CO_important}. 
However, accurate CO monitoring requires invasive catheters, which may lead to severe complications. 
Some previous deep learning works attempted to predict CO using the data of invasive medical devices \cite{adult_apco, asan2019apco, bibe2020}. 
However, we sought a non-invasive method. 
Our model allows us to monitor CO for general patients by eliminating the invasiveness.

The input variables of the regression task were the same as those of the hypotension prediction model. 
The output variable was stroke volume index (SVI) instead of CO so that we could return a prompt result and correct the interpatient biases. 
Note that SVI = CO/(heart rate (HR)$\times$body surface area).
To remove outliers, only values with CO / HR between 20 and 200mL/beat were used. 
The 20-s segments of input were extracted to predict immediate SVI values. 
Preprocessing for input segments was the same as the classification task.

\section{Experiments}
Training and testing datasets were obtained from VitalDB \cite{vitalrecorder}, an open-source physiological signal database containing perioperative physiological signs of more than 6,000 surgical patients.
We extracted the required tracks for each task and conducted minimal preprocessing to determine CNN models and attention mechanisms having the best model effects using real-world physiological signal data. 

To measure the effectiveness of the three attention mechanisms in each CNN model, performance variations were recorded by changing the attention fraction of the attention mechanism. 
Attention fraction is defined as the number of attention mechanisms divided by the number of CNN modules times 100.
The 0, 50, and 100\% attention fractions were considered in our experiments. 
Each attention mechanism was applied as the end-stage of each module. 
Note that for a 50\% attention fraction, one attention mechanism was embedded in every two CNN modules. 
Notably, the MSA-based model did not include a convolutional module and do not have a standardized architecture; hence, we explored various MSA-based model types using a grid search. 
The search spaces for self-attention had input and output dimensionalities of 16, 32, and 64, parallel attention layers (number of heads) of two, four, six, and eight, inner-layer dimensionalities of 32, 64, and 128, and identical layers (number of layers) of one, two, and three.
Through the hyperparameter search, we fixed other options as to be the best performance except number of heads and number of layers and recorded performance by changing the unit of number of heads and layers.
Note that unit for number of heads increase by 2 and for number of layers increase by 1.
The best setting of our MSA-based model was input and output dimensionality of 32, inner-layer dimensionality of 128.
A single convolutional layer was added to the input layer of each MSA-based model to match the variable dimensionality of the self-attention models. 

\begin{figure}
\includegraphics[width=\textwidth]{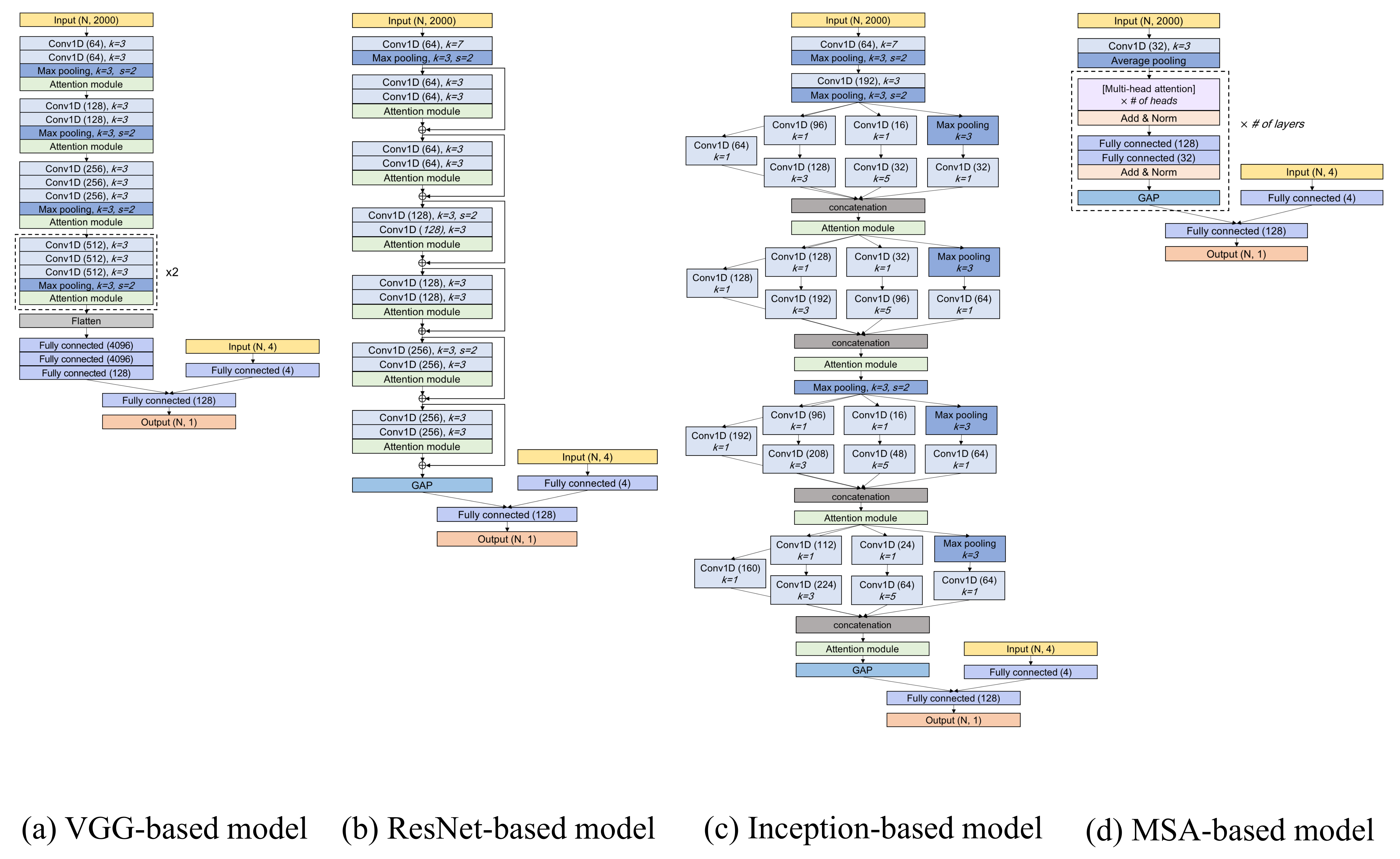}
\caption{The baseline model architectures. The models with 100\% attention fraction are shown. The number in parentheses means filters or neurons. \textit{k}: kernel size, \textit{s}: stride.}
\label{fig1}
\end{figure}

The input data of two tasks were two-channel (ECG and PPG) 100-Hz waveforms of 20-s. 
Patient demographic information was concatenated after the first fully connected layer. 
Detailed model architectures are illustrated in Fig \ref{fig1}. 
It presents the final baseline model used. 
The green box (attention module) was replaced with the module required for each experiment. 

For classification task, all models were trained with binary cross-entropy loss. 
The Adam optimizer was used for all models, apart from the inception-based one, which used RMSProp. 
The area under the receiver operating characteristics curve (AU-ROC) was used to evaluate the classification model. 
For the regression task, all models were trained with root mean squared error loss and the Adam optimizer. 
The mean absolute percentage error (MAPE) was calculated to measure model performance. 
Both classification and regression models were generated with a learning rate of 0.001 set to decrease by 0.1 times every 20 epochs. 
A batch size of 128 was used. 
To derive more reliable results, all models were repeated five times for training, and their performances were compared based on mean and standard deviation. 
We also measured the elapsed times of model convergence at given performances. 
The elapsed times to reach 0.7 AU-ROC for classification and 27.0\% of MAPE for regression task were considered.
All experiments, apart from those of the MSA-based models, were performed using Tensorflow 2.4.1 with Python 3.9 on a 32-core AMD EPYC 7542 processor and a single NVIDIA RTX 5000 GPU. 
For self-attention models, we used two NVIDIA RTX 5000 GPUs with NVLink connections to supplement GPU memory.

\section{Results}
Totals of 3,211 and 801 cases were extracted for hypotension and CO prediction, respectively. 
A randomly sampled 20\% of cases were used for testing.
For the hypotension prediction problem, 289,775 and 74,779 samples containing 4.74 and 4.03\% positive events were collected for training and testing, respectively. 
The CO prediction problem collected 271,288 and 64,659 samples, providing a mean SVI and a standard deviation of 42.11±13.25 and 41.71±12.37, respectively, for training and testing. 
Patient demographic information was not different (\textit{P}-value $>$ 0.05) between training and testing, except that the weight and height of patients in the hypotension testing were slightly larger (Table \ref{tab1}).

\begin{table}[ht]
\caption{Patient demographics of training and testing datasets}\label{tab1}
\begin{tabular}{|L{3cm}||R{3.5cm}|R{3.5cm}|R{1.6cm}|}
\hline
\multicolumn{4}{|c|}{\textbf{Hypotension prediction (Classification)}}\\
\hline
Characteristic &  Training dataset & Testing dataset & \textit{P}-value\\
\hline
Age, years\textsuperscript{$\dagger$} & 61.0 (49.0-69.8) & 60.0 (52.0-70.0) & 0.258\\
Sex, \# of male (\%) &  1409 (54.8\%)  & 368 (57.3\%) & 0.278\\
Height, cm\textsuperscript{$\dagger$} & 162.6 (156.3-168.7) & 163.4 (157.2-170.0) & 0.040\\
Weight, kg\textsuperscript{$\dagger$} & 60.0 (53.4-68.6) & 61.3 (53.0-68.3) & 0.030\\
\hhline{|=||=|=|=|}
\multicolumn{4}{|c|}{\textbf{Cardiac output prediction (Regression)}}\\
\hline
Characteristic &  Training dataset & Testing dataset & \textit{P}-value\\
\hline
Age, years\textsuperscript{$\dagger$} & 61.0 (52.0-70.0) & 62.0 (50.0-69.0) & 0.660\\
Sex, \# of male (\%) &  394 (61.3\%)  & 90 (57.0\%) & 0.367\\
Height, cm\textsuperscript{$\dagger$} & 163.8 (157.8-169.8) & 162.3 (155.4-169.2) & 0.178\\
Weight, kg\textsuperscript{$\dagger$} & 61.5 (54.2-69.5) & 61.1 (53.7-68.0) & 0.478\\
\hline
\multicolumn{4}{l}{\footnotesize{$\dagger$ Data are represented as median (interquartile range).}}
\end{tabular}
\end{table}

\begin{figure}
\includegraphics[width=\textwidth]{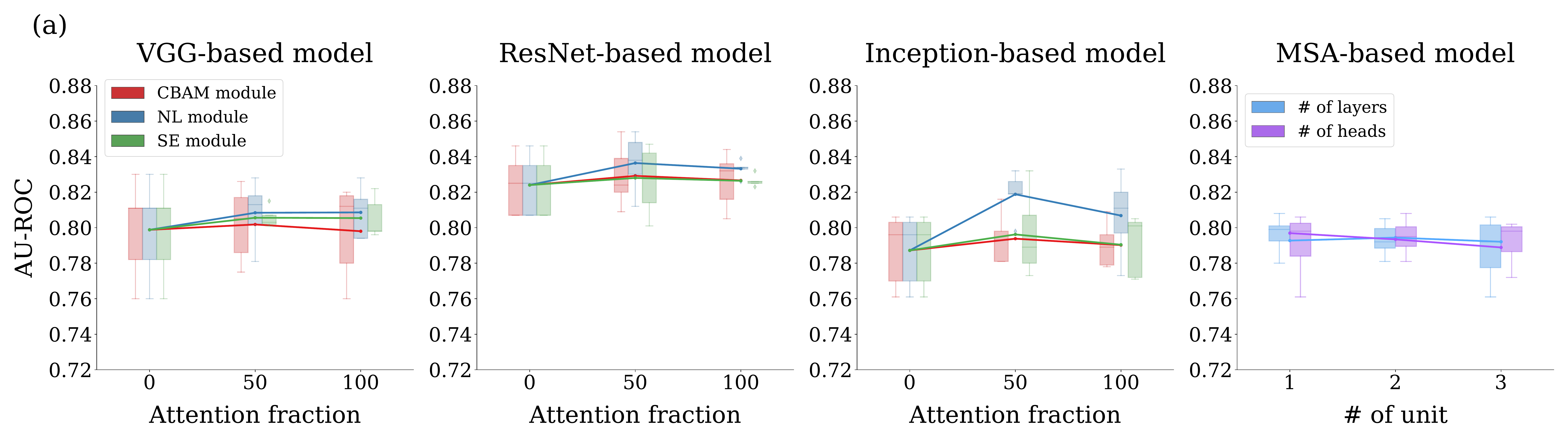}
\includegraphics[width=\textwidth]{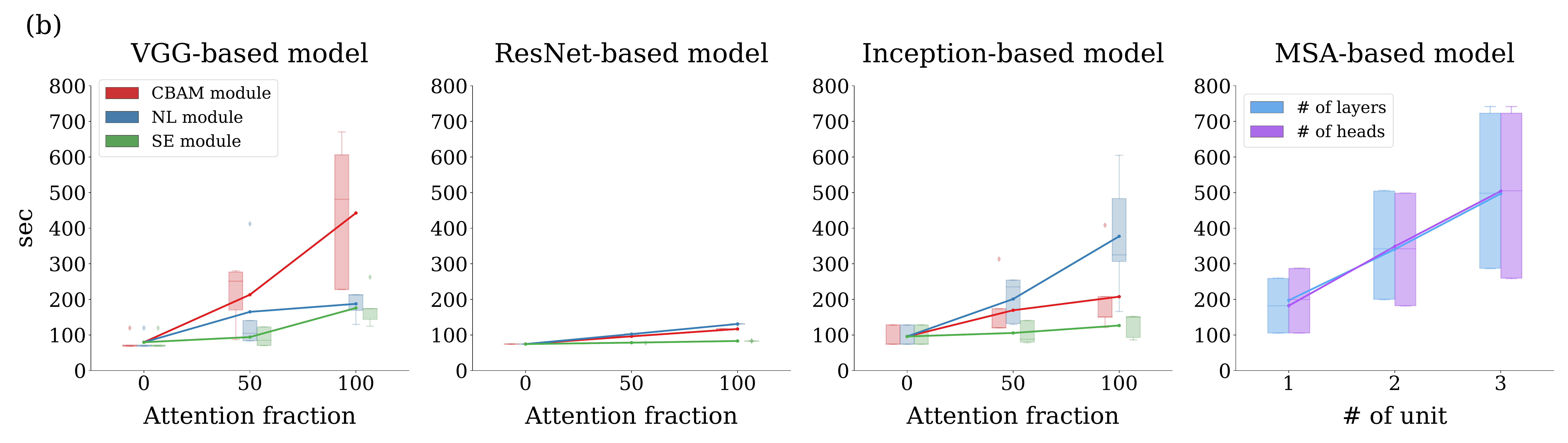}
\caption{Performance and convergence time in hypotension prediction problem. (a) is comparison of AU-ROC in the classification task. (b) is comparison of elapsed time to converge AU-ROC = 0.7} \label{fig2}
\end{figure}

\begin{figure}
\includegraphics[width=\textwidth]{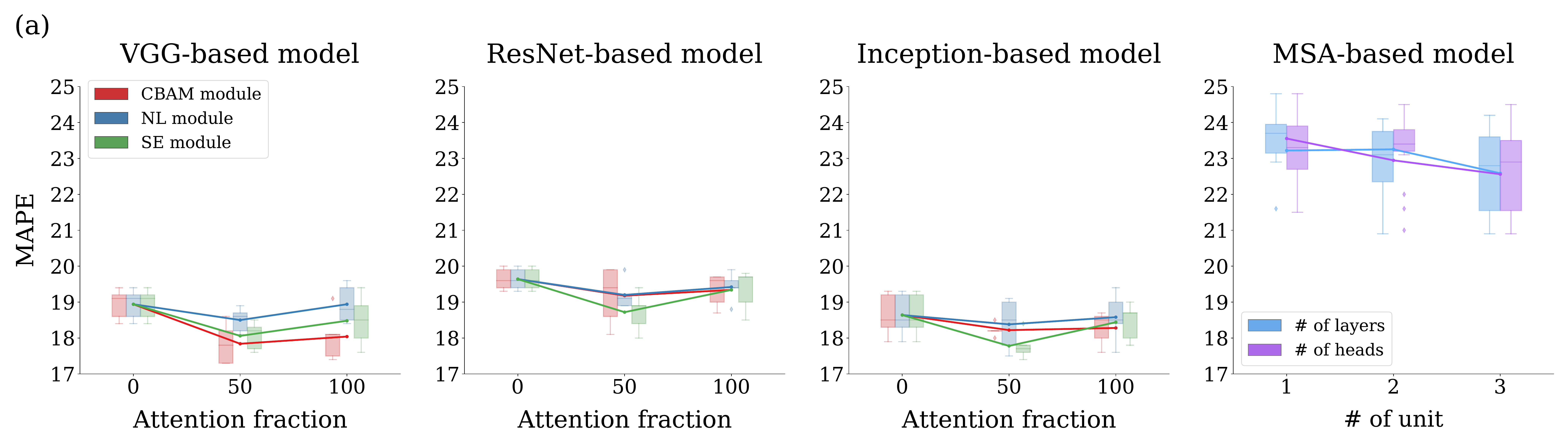}
\includegraphics[width=\textwidth]{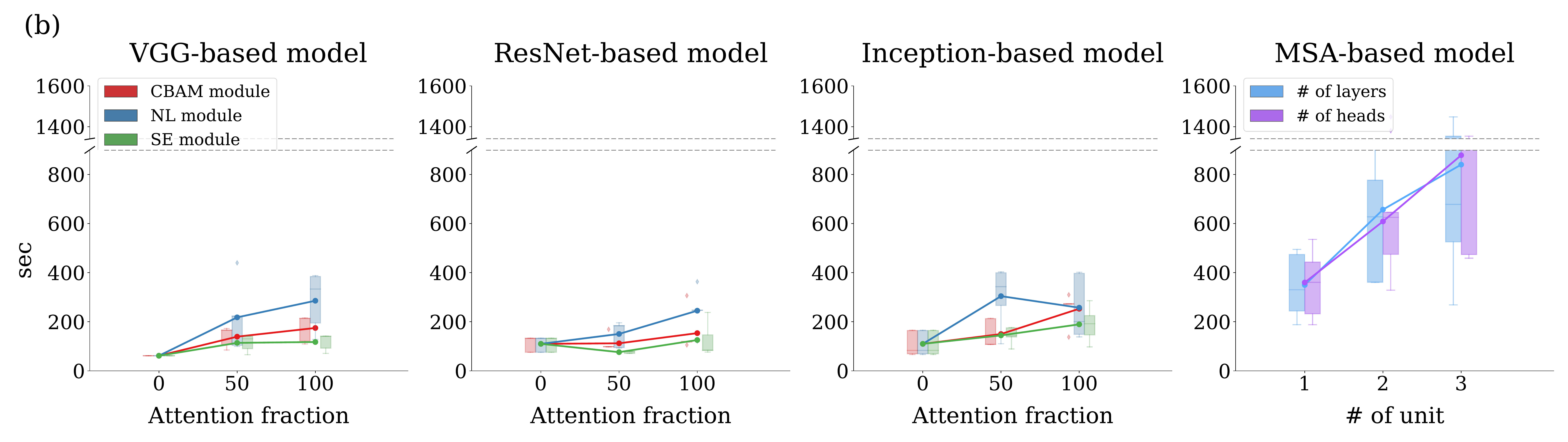}
\caption{Performance and convergence time in CO prediction problem. (a) is comparison of MAPE in the regression task. (b) is comparison of elapsed time to converge MAPE = 27.0\%} \label{fig3}
\end{figure}

The model performance variances of each CNN model and attention mechanism are illustrated in Fig \ref{fig2} and \ref{fig3}.
Regarding the classification task for predicting hypotension of Fig \ref{fig2}, ResNet-based model showed overall higher performance with a 50\% attention fraction. 
ResNet-based model with NL module showed the best AU-ROC of 0.854. 
When examining the elapsed time needed to converge 0.7 of the AU-ROC, ResNet-based model was the fastest.
Additionally, the SE module added negligible additional computing overhead, but the overall CNN performance increased. 
There was an obvious tendency of increased performance when using \textit{spatial} attention (i.e., NL or CBAM module). 

During CO regression prediction, as shown in Fig \ref{fig3}, the VGG-based model showed an overall low error. 
The VGG-based model with a 50\% attention fraction of the CBAM module showed the best MAPE of 17.3\%.
However, ResNet-based model had the best convergence time to achieve 27.0\% of MAPE. 
The computational overhead of the SE module in the three CNN models was also negligible in the regression problem, whereas it played a major role in reducing errors.
Moreover, the convergence time was shortened in the ResNet-based model with SE module.
There was also a clear tendency of decreasing error when using \textit{channel} attention mechanisms (i.e., SE or CBAM module).

These experimental results can be better understood when contrasted with the problem defined. 
To predict hypotension within 5 min of occurrence, the most important feature is hemodynamic flow changes across 20-s of input data. 
Therefore, \textit{spatial} attention plays an important role in model performance. 
In the prompt-CO regression problem, the waveform shape from a single beat was most important as CO is closely related to heart dynamics and the elasticity or compensation of blood vessels.
Notably, each patient has a different beat pattern. 
Therefore, it is crucial to properly analyze the shape of the beat waveform.
\textit{Channel} attention extracts various features from the input and improves performance by helping diversify feature representations.

In both problems, the model performance was generally better when using 50\% of the attention fraction rather than 100\% or the fully self-attention-based model. 
Similar results were reported for computer vision problems\cite{howdoVITwork}.
We confirmed that convolution and self-attention were complementary in physiological signal deep learning, as with computer vision. 
Furthermore, good performance cannot be achieved by using only one building block.

\section{Conclusion}

In this study, we determined the best CNN and attention mechanism pairing for building deep learning models for physiological signal analysis. 
An attention mechanism should be selected by determining which characteristics from the raw physiological signal should be addressed to solve the problem. Convolution and attention mechanisms are complementary; therefore, there may be an ideal attention fraction for optimal performance. 
The ResNet-based model showed moderate performance and fast convergence in both experimental tasks. 
Therefore, ResNet-based model with an attention mechanism is the best candidate for prototype model. 
Recent studies suggest using a combined MSA with CNN for higher performance.
We plan to compare physiological signal analysis performance using multiple models in a future paper.

\subsubsection{Acknowledgement.}
This research was supported by a grant of the Korea Health Technology R\&D Project through the Korea Health Industry Development Institute (KHIDI), 
funded by the Ministry of Health \& Welfare, Republic of Korea (grant number : HI21C1074); 
and the Korea Medical Device Development Fund grant funded by the Korea government 
(the Ministry of Science and ICT, the Ministry of Trade, Industry and Energy, the Ministry of Health \& Welfare, Republic of Korea, and the Ministry of Food and Drug Safety) 
(Project Number: 202011B23)

%
%
\bibliographystyle{splncs04}
\bibliography{paper1798.bib}
\newpage
\newpage
\appendix
\section{Appendix}
The original CNN models (e.g., VGG-16, ResNet-18, and Inception-V1) used 224 × 224 image inputs and analyzed them with two-dimensional (2D) convolutions. 
However, in our research, the dimensionality of input data should be one-dimensional (1D). 
Therefore, all 2D convolutional operations were replaced with 1D convolutions. 
Additionally, input sizes were much smaller at 224 × 224 = 50,176 vs. 2,000. 
We thus reduced the model depth to prevent overfitting caused by superfluous immoderate trainable parameters.
Let our input data size of 2,000 to be 2D. 
2,000 $\approx$ 45 × 45. 
Thus, the ratio between image data used in the original CNN studies and our physiological data was 224 / 45 $\approx$ 5. 
Therefore, we used the model reduction ratio of five for each CNN model.
The main characteristics of CNN models was the modules they contained. 
The VGG module included two or three consecutive convolution layers and a pooling layer. 
The ResNet module contained two consecutive convolution layers and a residual path. 
The inception module included three convolution layers and a pooling layer in parallel. 
To maintain each model’s identity, we set cutoff criteria while preserving the modules. 
Fig \ref{figA} illustrates the shallow part of each CNN model. 

\begin{figure}
\setcounter{figure}{0}
\renewcommand{\figurename}{Fig.}
\renewcommand{\thefigure}{A\arabic{figure}}
\includegraphics[width=\textwidth]{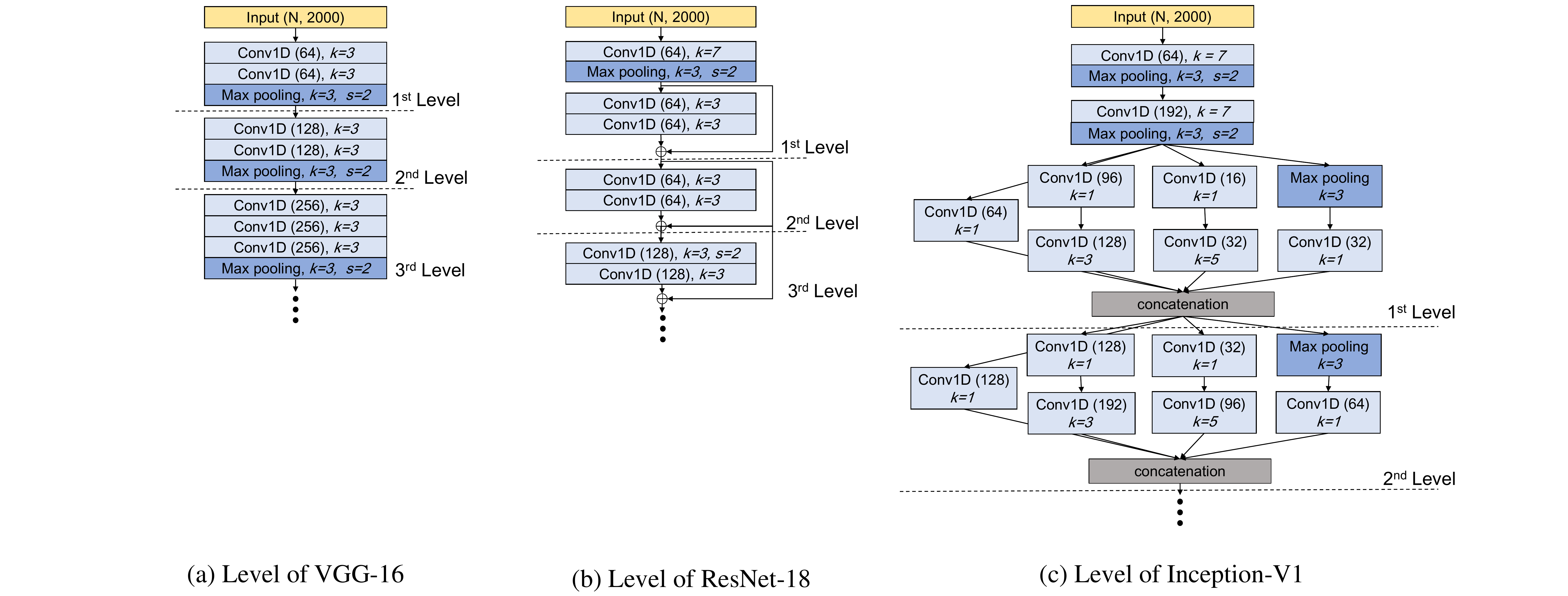}
\caption{Example of the levels of each model} \label{figA}
\end{figure}

\begin{table}[ht]
\setcounter{table}{0}
\renewcommand{\thetable}{A\arabic{table}}
\caption{Trainable parameters for each level of each model}\label{tabA1}
\begin{tabular}{|C{1.5cm}||C{3.3cm}|C{3.3cm}|C{3.5cm}|}
\hline
Level &  VGG-based model Trainable param. & ResNet-based model Trainable param. & Inception-based model Trainable param.\\
\hline
1 & 192,128,065 & 26,048 & 117,744\\
2 & 189,891,329 & 51,008 & 297,584\\
3 & 130,614,785 & 134,080 & 538,592\\
4 & 90,639,361 & 233,152 & \bf806,504\\
5 & \bf40,567,296 & 563,136 & 1,089,280\\
6 & - & \bf 957,888 & 1,404,864\\
7 & - & 2,273,216 & 1,884,096\\
8 & - & 3,849,152 & 2,538,432\\
\hhline{|-||-|-|-|}
Default param. & 40,567,296 & 3,849,152 & 3,417,264\\
Default/5 param. & 8,133,459 & 769,830 & 683,453\\
\hline
\end{tabular}
\end{table}

Note that the level indicates a section divided while maintaining the module’s property.
To find the optimal subset of the CNN model for our study, we compared the number of training parameters by dividing the model by levels. 
The fully connected part (the classification or regression part) of the original model was added to the subset model. 
Additional concatenating layers for patient demographic data were added in the last fully connected part. 
Table \ref{tabA1} presents the number of trainable parameters divided by the level of each model. 
ResNet and Inception models showed fewer trainable parameters as they were divided at shallow levels, whereas VGG showed more parameter increases owing to the growth of feature sizes entering the fully connected layer without global-average pooling. 
We selected a VGG five levels (full model), a ResNet six levels, and an inception with four levels as our baseline CNN architecture.

\end{document}